\documentclass[floats,preprint,prl,aps]{revtex4}
\usepackage{latexsym}
\usepackage{graphicx}
\newcommand\beq{\begin{equation}}
\newcommand\eeq{\end{equation}}
\newcommand\bea{\begin{eqnarray}}
\newcommand\eea{\end{eqnarray}}
\newcommand\non{\nonumber}
\newcommand\bib{\bibitem}

\begin{document}
\title{\bf Emergent of Majorana Fermion mode and Dirac Equation in Cavity 
Quantum Electrodynamics  
}

\author{\bf Sujit Sarkar}
\address{\it Poornaprajna Institute of Scientific Research,
4 Sadashivanagar, Bangalore 5600 80, India.\\
}
\date{\today}

\begin{abstract}
We present the results of low lying excitation of coupled optical
cavity arrays. We derive the Dirac equation for this system  
and explain the existence of Majorana fermion mode in the system.
We present quite a few analytical relations
between the Rabi frequency oscillation and the atom-photon
coupling strength to achieve the different physical situation
of our study and also the condition for massless excitation in the system. 
We present several analytical relations between
the Dirac spinor field, order and disorder operators for our
systems. We also show that the Luttinger liquid physics is one 
of the intrinsic concept in our system.   
\end{abstract}
\maketitle

{\bf Introduction:}  
The recent experimental success in engineering strong interaction 
between the photons and atoms in high quality micro-cavities opens up
the possibility to use light matter system as quantum simulators for
many body physics [1-21]. 
Many interesting results are coming out to understand the complicated 
quantum many body system. The Bose-Hubbard model, quantum spin model
and the other exotic quantum phases of the quantum many body system
have already been studied [3-7]. 
The further application of the basic principle
of cavity-QED system is the circuit QED [22-26]. When a qubit (qubits) coupled to
the high quality LC circuits, it presents the same physical picture with few
extra achievement over the conventional cavity-QED system. A focus
on the coupled cavities is one of the most potential candidate for
an efficient quantum simulator due to the control of the microcavities
parameters and success of fabrication of large scale cavity arrays [25-26].
This is the very brief discussion of the presence status of the cavity QED
system.\\
In the present study one of our goal is to predict the presence
of Majorana fermions in our model system. Before we proceed further
, we would like to describe very briefly about the appreance of
Majorana fermions in quantum condensed matter system.
Majorana had introduced a special kind of fermions which are their
own antiparticle, i.e., the neutral particle \cite{majo,wil}. He had introduced
this particle to describe neutrions. In recent years, there are
several candidates of Majorana fermions in quantum condensed
matter system like quantum Hall system with filling fraction 
$ 5/2 $ \cite{read,read2}. Kitaev at first found the existence of Majorana 
fermion mode in one dimensional model \cite{kitaev}. Many research 
group have already been proposed the physically existence of
MFs at the edge state of 1D system like electrostatic defects lines
in superconductor, quasi-one dimensional superconductor and
cold atom trapped in one dimension \cite{wimmer,kop}.
Majorana fermions are obey non-Abelian statics both in 2D and
1D, allowing of certain gate operation required in quantum
computation \cite{nayak}. 
Due to the non-local character, the qubit built
out of Majorana fermions are insensitive to local parity 
conserving perturbation \cite{nayak,lut,oreg,potter,jiang}. 
The search for experimentally accessible systems that are described
by a Dirac equation has received much attention in recent years 
\cite{nayak,lut,oreg,potter,jiang}.
In this research paper, we present an extensive derivation of Dirac equation
and also the existence of Majorana fermions mode in an optical
cavity array. We also present the analytical
relation between the Rabi frequency oscillation and the atom-photon
coupling strengths to mimic the transverse Ising model, Dirac equation,
magnetic ordered state, quantum paramagnetic state and massless excitation.
Quantum state engineering of the optical cavity array system is
in the state of art due to the rapid technical development of this 
field [1] therefore one can achieve these quantum phases in the
laboratory.\\ 
{\bf The Model Hamiltonian and Majorana Fermion Modes:\\} 
The Hamiltonian of our present study consists of three parts:
\beq
H ~= ~ {H_A} ~+~ {H_C}~+~{H_{AC}}
\eeq     
The Hamiltonians are the following
\beq
{H_A} ~=~ \sum_{j=1}^{N} { {\omega}_e } |e_j > <e_j | ~+~ 
{\omega}_{ab} |b_j > <b_j | 
\eeq
where $j$ is the cavity index. ${\omega}_{ab} $ and ${\omega}_{e} $ are 
the energies of the state $ | b> $ and the excited state respectively. The
energy level of state $ |a > $ is set as zero. $|a>$ and $|b> $ are
the two stable state of a atom in the cavity and $|e> $ is the
excited state of that atom in the same cavity. 
The following Hamiltonian describes the photons in the cavity,
\beq
 {H_C} ~=~ {{\omega}_C} \sum_{j=1}^{N} {{a_j}}^{\dagger} {a_j} ~+~
{J_C} \sum_{j=1}^{N} ({{a_j}}^{\dagger} {a_{j+1}} + h.c ),  
\eeq 
where ${a_j}^{\dagger}({a_j})$ is the photon
creation (annihilation) operator for the photon field in the $j $'th cavity, ${\omega}_C $
is the energy of photons and $ J_C $ is the tunneling rate of photons
between neighboring cavities.
The interaction between the atoms and photons and also by the driving lasers
are described by
\beq
{H_{AC}}~=~ \sum_{j=1}^{N} [ (\frac{{\Omega}_a}{2} e^{-i {{\omega}_a} t} +
{g_a} {a_j}) |e_j > < a_j | + h.c] + [a \leftrightarrow b ] . 
\eeq
Here ${g_a} $ and ${g_b} $ are the couplings of the cavity mode for the
transition from the energy states $ |a > $ and $ | b> $ to the excited state.
${\Omega}_a $ and ${\Omega}_b $ are the Rabi frequencies of the lasers
with frequencies ${\omega}_a $ and $ {\omega}_b $ respectively.\\
The authors of Ref. \cite{hart1,hart2,sujop}
have derived an effective spin model by considering the following physical
processes:
A virtual process regarding emission and absorption of
photons between the two stable  states of neighboring cavity yields the resulting 
effective Hamiltonian as
\beq
{H_{xy}} = \sum_{j=1}^{N}  B {{\sigma}_j}^{z} ~+~\sum_{j=1}^{N} 
(\frac{J_1}{2} {{\sigma}_j}^{\dagger} {{\sigma}_{j+1}}^{-} ~+~
\frac{J_2}{2} {{\sigma}_j}^{-} {{\sigma}_{j+1}}^{-} + h.c )
\eeq 
When $J_2 $ is real then this Hamiltonian reduces to the XY model.
Where ${{\sigma}_j}^{z} = |b_j > <b_j | ~-~ |a_j > <a_j | $,
${{\sigma}_j}^{+} = |b_j > <a_j | $, ${{\sigma}_j}^{-} = |a_j > <b_j | $
. 
\bea
H_{xy} & = & \sum_{i=1}^{N} ( B  {{\sigma}_i}^{z}~+~ {J_1} ( {{\sigma}_i}^{x}
{{\sigma}_{i+1} }^{x}  + {{\sigma}_i}^{y}
{{\sigma}_{i+1} }^{y} ) \non\\
& & + {J_2} ( {{\sigma}_i}^{x}
{{\sigma}_{i+1} }^{x} - {{\sigma}_i}^{y}
{{\sigma}_{i+1} }^{y} ) ) \non\\    
& & =  \sum_{i=1}^{N} B ( {{\sigma}_i}^{z}~+~{J_x} {{\sigma}_i}^{x}
{{\sigma}_{i+1}}^{x} ~+~ {J_y} {{\sigma}_i}^{y}
{{\sigma}_{i+1}}^{y}) .
\eea
With ${J_x} = (J_1 + J_2 ) $ and ${J_y} = (J_1 - J_2 ) $.\\
We follow the references \cite{james,hart1}, to present the analytical 
expression for the different physical parameters of the system.\\
\beq
B = \frac{\delta_1}{2} - \beta 
\eeq
\bea
\beta  & = &  \frac{1}{2} [\frac{{|{\Omega_b}|}^2 }{4 {\Delta}_b }
({\Delta}_b - \frac{{|{\Omega_b}|}^2 }{4 {\Delta}_b } -  
 \frac{{|{\Omega_b}|}^2 }{4 ( {\Delta}_a  - {\Delta}_b )} - {\gamma_b} {g_b}^2 
- {\gamma_1} {g_a}^2 \non\\ 
& & + {\gamma_1}^2 \frac{{g_a}^4 }{{\Delta_b}} - (a \leftrightarrow b)] 
\eea
\beq
{J_1} = \frac{\gamma_2}{4} ( \frac{{|{\Omega_a}|}^2 {g_b}^2 }{{ {\Delta}_a }^2 }
 +  \frac{{|{\Omega_b}|}^2 {g_a}^2 }{{ {\Delta}_b }^2 } ) ,
{J_2} = \frac{\gamma_2}{2} ( \frac{{\Omega_a} {\Omega_b} g_a g_b }{{\Delta}_a {\Delta_b} }
 ).
\eeq
Where 
$ \gamma_{a,b} = \frac{1}{N} \sum_{k} \frac{1}{ {\omega}_{a,b} - {\omega}_k } $
$ \gamma_{1} = \frac{1}{N} \sum_{k} \frac{1}{ ( {\omega}_{a}+  {\omega}_{b})/2 - {\omega}_k } $ and
$ \gamma_{2} = \frac{1}{N} \sum_{k} \frac{e^{ik} }{ ( {\omega}_{a}+  {\omega}_{b})/2 - {\omega}_k } $
${\delta_1} = {\omega}_{ab} - ({\omega}_a - {\omega}_b )/2 $, 
${\Delta}_a = {\omega}_e - {\omega}_a$.  
${\Delta}_b = {\omega}_e - {\omega}_a -({\omega}_{ab} - {\delta_1})$.
${{\delta}_a}^{k} = {\omega}_e - {\omega}_k $,
${{\delta}_b}^{k} = {\omega}_e - {\omega}_k  -({\omega}_{ab} - {\delta_1}) $,
${\omega}_k = {\omega}_c + J_c \sum_{k} cosk $.
$g_a$ and $g_b$ are the couplings of respective transition to the cavity mode,
${\Omega}_a $ and ${\Omega}_b$ are the Rabi frequency of laser with frequency
$\omega_a $ and $\omega_b $. \\
The system reduces to Ising model with transverse field at $ J_1 = J_2 $,
i.e., $J_x $ become $ J_1 + J_2 $ and  $ J_y =0 $. The effective Hamiltonian become
the transverse Ising model which studied in the previous literature \cite{ss,mussardo,druff}. 
Here our main
motivation is to use some of important results of this model Hamiltonian 
to discuss the relevant physics of array of cavity QED system.\\
Before we proceed further, we would like to discuss in detail  
the analytical relation between the different
coupling constants of cavity QED system to achieve this Hamiltonian.
In the microcavity array, the condition for $J_1 = J_2 $ achieve when
\beq
{{\Omega}_a}^2  {g_b}^2 {\Delta_b}^2 + {{\Omega}_b}^2  {g_a}^2 {\Delta_a}^2 =
2 {\Omega}_a {\Omega}_b  g_a g_b {\Delta}_a {\Delta}_b . 
\eeq
The above condition implies that $ {\Omega}_a = {\Omega}_b  \frac{g_a \Delta_a}{g_b \Delta_b }$.
The only constraint is that ${\Delta}_a \neq {\Delta}_b $, the magnetic field 
diverge when ${\Delta}_a = {\Delta}_b $. At the same time, ${\Omega}_a = {\Omega}_b $ and
$g_a = g_b$ are also not possible because this limit also leads to the condition
${\Delta}_a = {\Delta_b} $. Suppose we consider,
${\Omega}_a  = \alpha_1 {\Omega}_b $, $g_a = \alpha_2 g_b $ and ${\Delta}_a = \alpha_3 {\Delta}_b $.
These relations implies that 
$ {\alpha_1 }^2 + {\alpha_2 }^2 {\alpha_3 }^2 = 2 {\alpha}_1 {\alpha}_2 {\alpha}_3 $.
${\alpha}_1 = {\alpha}_2  {\alpha}_3  $, 
${\alpha}_1 , {\alpha_2 }$ and ${\alpha_3}$ are the
numbers. These analytical relations help to implement 
the transverse Ising model Hamiltonian but $\alpha_1 $, $\alpha_2 $ and $\alpha_3 $ should
not be equal to 1.\\
The quantum state engineering of cavity QED is in the state of art due to the
rapid progress of technological development of this field [1]. Therefore one can 
achieve this limit to get the desire quantum state. 
\beq
H_{T} = B \sum_{j=1}^N ( {{\sigma}_z} (j) + \lambda {{\sigma}_x} (j) {{\sigma}_{x}} (j+1) ),
\eeq
where $\lambda = \frac{ J_1 + J_2}{B} $. The transverse Ising model was studied widely
in the literature and also exhibit a quantum phase transition between the
magnetically ordered state to the quantum paramagnetic phase for $\lambda > 1$
and $\lambda < 1$ respectively \cite{ss,mussardo,druff}.\\
Now we express the condition for the magnetic order phase and quantum paramagnetic
phase in terms of the physical parameters of the optical cavity QED system which
gives us the relevant physics of the system.\\
The condition for the magnetic ordered system can be expressed as
\beq
\frac{\gamma_2 }{4} ( \frac{{\Omega}_a {g_b}^2 }{{\Delta_a }^2 }  
+  \frac{{\Omega}_b {g_a}^2 }{{\Delta_b }^2 } ) + 
\frac{\gamma_2}{2} ( \frac{ \Omega_a \Omega_b g_a g_b}{ \Delta_a \Delta_b } ) 
> \omega_{ab} - \frac{ \omega_a - \omega_b}{2} -2 \beta  
\eeq
The condition for the quantum paramagnetic phase is \\
\beq
\frac{\gamma_2 }{4} ( \frac{{\Omega}_a {g_b}^2 }{{\Delta_a }^2 }
+  \frac{{\Omega}_b {g_a}^2 }{{\Delta_b }^2 } ) +
\frac{\gamma_2}{2} ( \frac{ \Omega_a \Omega_b g_a g_b}{ \Delta_a \Delta_b } )
< \omega_{ab} - \frac{ \omega_a - \omega_b}{2} -2 \beta
\eeq
When the applied magnetic field is absent, 
the effective Ising model has two degenerate ground states. The ground states are
$ |A > = \Pi_{j} {|\rightarrow >}_j $, $ |B > = \Pi_{j} {|\leftarrow >}_j $.
For a finite magnetic field but less than $ J_1 + J_2 $, the system has a tendency 
to flip the pseudo spin. 
At that phase one can write down the true eigen state,
$ |{\psi}_A > = \frac{1}{\sqrt{2}} ( |A> + |B >) $ ,
$ |{\psi}_B > = \frac{1}{\sqrt{2}} ( |A> - |B >) $. 
Now our main intention is to recast this spin model in spinless
fermion model through Jordon-Wigner transformation which relate the
spin operators to the spinless fermion operators. 
We use the following relation:\\
${\sigma}_z = 2 c^{\dagger}(j) c(j) - 1 $,
$ {\sigma}_x (j) {\sigma}_x (j+1) = ( c^{\dagger} (n) - c (n) ) ( c^{\dagger} (n+1) - c (n+1) )$.\\
One can write the Hamiltonian after the Jordon-Wigner transformation as
\beq
H = 2 \sum_{j= 1}^{N} {c}^{\dagger} (j) c(j) + \lambda ( c^{\dagger} (j)
- c (j) ) (c^{\dagger} (j+ 1)
- c (j+1) )
\eeq
We solve this Hamiltonian, to get the energy spectrum by taking the Fourier
transform.\\
$ c(j) = \frac{1}{\sqrt{N}} \sum_{k} e^{-i k a} $,
$ c^{\dagger} (j) = \frac{1}{\sqrt{N}} \sum_{k} e^{i k a} $.
Where $c_k $ and ${c_k}^{\dagger} $ are the fermionic annihilation and
creation operator in momentum space.\\ 
The Hamiltonian reduce to
\bea
H & = & 2 \sum_{k> 0} (1 + \lambda cosk) 
({c_k}^{\dagger} {c_k} + {c_{-k}}^{\dagger} {c_{-k}}) \non\\
& & + 2 i \lambda  \sum_{k > 0} sink  ({c_k}^{\dagger} {c_{-k}}^{\dagger} + 
{c_{k}} {c_{-k}})
\eea  
Now our main task is to express the Hamiltonian in the diagonalized form. We
follow the Bogoliubov transformation.\\
$ \eta_{k} = \alpha_k c_k  + i {\beta}_k {c_{-k}}^{\dagger}$  
and $ \eta_{-k} = \alpha_k c_{-k}  - i {\beta}_k {c_k}^{\dagger}$, $k>0$.\\
The operator ${\eta}_{k} $ and $ {{\eta}_k}^{\dagger} $ are the fermionic operators.
We use the following relations,\\ 
$ \{ {\eta}_k , {{\eta}_p}^{\dagger} \} = \delta_{k,p} $,
$ \{ {\eta}_k , {{\eta}_p} \} = 0 $,
$ \{ {{\eta}_k}^{\dagger} , {{\eta}_p}^{\dagger} \} = 0 $.
This relation implies, $ {{\alpha}_k}^{2}  +  {{\beta}_k}^{2} =1 $.
One can also revert the relation between $c_k$ and ${\eta}_k $. We also 
parameterize $\alpha_k = cos \theta_k $ and $\beta_k = sin \theta_k $.
One can express the transformed Hamiltonian in two parts
\beq
 H = H_A  + H_B 
\eeq
\bea
H_A  & = & \sum_{k>0} [-2 (1 + \lambda cosk) ( {{\alpha}_k}^2 - {{\beta}_k}^2 )
+ 4 \lambda sink {\alpha}_k {\beta}_k ] \non\\  
& &  ({{\eta}_k}^{\dagger} {\eta_k} 
 {{\eta}_{-k}}^{\dagger} {\eta_{-k}} ) 
\eea
\bea 
H_B  & = &  \sum_{k>0} [ 4 i (1 + \lambda cosk ) {\alpha_k} {\beta_k} + 2 i \lambda
sink ( {{\alpha}_k}^2 - {\beta_k}^2 )] \non\\ 
& & ({{\eta}_k}^{\dagger} {\eta_{-k}}^{\dagger}
{{\eta}_{k}} {\eta_{-k}} ) 
\eea
To express this Hamiltonian in the diagonal form, we find the following relation\\
$ 4 (B + \frac{\gamma_2 {\Omega_b }^2 {g_a}^2 }{ {\Delta_b}^2 } cosk ) {\alpha_k} {\beta_k} + 
2 \frac{\gamma_2 {\Omega_b }^2 {g_a}^2 }{ {\Delta_b}^2 } sink 
({\alpha_k}^2 - {\beta_k}^2) = 0 $.\\ 
Finally this gives the condition,
\beq
 tan 2 {\theta_k} 
 = \frac{2 \gamma_2 {\Omega_b }^2 {g_a }^2 }{ 
 2 {\gamma_2} {\Omega_b}^2 {g_a}^2  cosk - ({\delta_1} - 2 \beta ) {\Delta_b}^2 } 
\eeq 
$ 2 {\alpha}_k {\beta}_k = sin2 {\theta_k} $,
$ {\alpha_k}^2 - {\beta_k}^2 = cos 2 \theta_k $.
Now we analysis the spectrum: \\
\beq
H = 2 \sum_k \Omega_{K} {\eta_k}^{\dagger} {\eta_k}
\eeq
We can express this energy spectrum in terms of Rabi frequency
oscillation, atom-photon coupling strength,\\
\beq
\Omega_k  =  \frac{2}{{\delta_1}/2 -\beta} \sqrt{ {({\delta_1}/2 -\beta )}^2 
+ {\gamma_2 }^2 \frac{ {{\Omega_b}}^4 {g_a }^4 }{ {\Delta_a }^4 }  
+ 2 ( {\delta_1}/2 - \beta ) {\gamma_2 } \frac{ {{\Omega_b}}^2 {g_a }^2 }{ {\Delta_b }^2 } } 
\eeq
The minimum occurs at $ k = \pm \pi$, 
$\Omega_{k = \pm \pi} = 2 |1 - \lambda | =2 |1 - \frac{2 \gamma_2}{\delta_1 - 2 \beta}
\frac{ {{\Omega}_b}^2 {g_a}^2}{{\Delta_b}^2 }|$. 
We are interested in the continuum limit
and also restore the lattice spacing $\alpha$ and measure the momentum
w.r.t the minimum value,
$ k = \pi + k^{'} \alpha $ . The energy expression which contains the physical
dimension of energy is $ E( k^{'} ) =  \frac{\Omega_k}{2 \alpha} $.
In this limit, 
$ E( k^{'} ) = \sqrt{ {(\frac{1 - \lambda}{\alpha})}^2  + \lambda {k^{'} }^2 } $.
If $\lambda$ is
close to a critical value $\lambda \sim 1 $, we then have the dispersion of a
particle with mass $ m = \frac{ 1-\lambda}{\alpha} $.\\
If $\lambda =1$, it becomes the massless particle $ E (k^{'} ) \sim k^{'} $.\\
In the cavity QED system, we can express the condition for massless excitation of the
system as $ ({{\delta_1}/2 -\beta}){\Delta_b}^2 = {\gamma_2} {\Omega_b}^2 {g_a}^2 $.
In terms of fields, $ \eta (a) = \frac{1}{\sqrt{N}} \sum_k e^{i ka} {\eta}_k $.
$ {\eta (a)}^{\dagger} = \frac{1}{\sqrt{N}} \sum_k e^{- i ka} {{\eta}_k}^{\dagger} $.\\
$ {\chi}_1 (a) = (1/2) ( \eta (a) + {\eta (a)}^{\dagger})$,
$ {\chi}_2 (a) = (1/2i) ( \eta (a) - {\eta (a)}^{\dagger})$ \\
$ {{\chi}_1 (a)}^{\dagger} = {\chi}_1 (a) $, $ {{\chi}_2 (a)}^{\dagger} = {\chi}_2 (a) $. \\
$ \{ {\chi_1} (x1), {\chi_2} (x2) \} = {\delta}_{x1,x2} {\delta}_{1,2} $. 
Therefore, ${\chi}_{1} $ and ${\chi}_{2} $ are satisfying the all properties of neutral
fermionic fields what Majorana proposed.\\
The authors of \cite{choi2} have investigated the low lying excitation
of one dimensional array of circuit QED (cktQED) with each
cktQED being in the ultra strong coupling regime and they
have found the Majorana bound state. But the
starting Hamiltonian  
of our system is completely different.\\
Now we calculate the energy density of the system using Eq. 21. We would like
to integrate the dispersion spectrum ${\Omega}_k $ to get the energy
density. The analytical expression for energy density is
\beq
{\epsilon}_0  = \frac{2}{\pi}~(1 +\frac{2 \gamma_2}{\delta_1 - 2 \beta}
\frac{ {{\Omega}_b}^2 {g_a}^2}{{\Delta_b}^2 } )~ E (\frac{\pi}{2}, \sqrt{1-{\gamma}^2 } ) 
\eeq 
Where $ {\gamma}= |\frac{ (\delta_1 -2 \beta){\Delta_b}^2 - 2 {\gamma_2} {\Omega_b}^2 {g_a}^2 }
{ (\delta_1 -2 \beta){\Delta_b}^2 + 2 {\gamma_2} {\Omega_b}^2 {g_a}^2 }|$,
~ $ E (\frac{\pi}{2}, \sqrt{1-{\gamma}^2 })$ is the complete Elliptic integral
of 2nd kind. 
After a little bit of calculation, we obtain
the analytical expression in the asympototic limit for energy density.
\bea
{\epsilon}_0 & = & 1 + 1/2 ~ ( ln| 4 \frac{ (\delta_1 -2 \beta){\Delta_b}^2 + 2 {\gamma_2} {\Omega_b}^2 {g_a}^2 }
{ (\delta_1 -2 \beta){\Delta_b}^2 - 2 {\gamma_2} {\Omega_b}^2 {g_a}^2 }| - 1/2 ) {\gamma}^2  \non \\ 
& & + 3/16  ( ln| 4 \frac{ (\delta_1 -2 \beta){\Delta_b}^2 + 2 {\gamma_2} {\Omega_b}^2 {g_a}^2 }
{ (\delta_1 -2 \beta){\Delta_b}^2 - 2 {\gamma_2} {\Omega_b}^2 {g_a}^2 }| - 13/12 ) {\gamma}^4 
\eea
 
{\bf Derivation of Dirac Equation and Condition of Massless Excitations:}\\
During the derivation of Dirac equation we rotate the y-axis of the spin basis
through $\pi/2 $. Through this rotation z-axis become x-axis and x-axis become -z-axis.
The physics of the system remain the same.
The starting Hamiltonian of our system now becomes\\
\beq
H = \sum_{n} [\lambda s_z (n) s_z (n+1) + s_x (n)].
\eeq
We recast the Hamiltonian in the following form because it will help us to
use the order and disorder operator directly to the derivation of the equation
of motion and finally the Dirac equation. Here we present an extensive derivation
of Dirac equation for our model system.\\
Kramers-Wannier symmetry for two dimensional Ising model reveals for
the case of one-dimensional quantum Ising chain through the dual 
lattice to the original one (site index n). 
Here we introduce the order and disorder operator following \cite{mussardo,druff}.
These operators are defining the sites of the dual lattice, i.e., we
define the operator between the nearest-neighbor site of the original
lattice. The analytical relation between the Pauli operators and $\mu $
operators are the following:\\
\beq
 {{\mu}_z }^2 = 1 = {{\mu}_x }^2 ,
\eeq
\beq
{{\mu}_z } (n -1/2 ) {{\mu}_z } (n +1/2 ) = {\sigma}_x (n).
\eeq
\beq
{{\mu}_x } (n + 1/2) = {\sigma}_z (n) {\sigma}_z (n+1),
\eeq
\beq
{{\mu}_z } (n + 1/2) = \Pi_{j=1}^{n}  {\sigma_x} (j).
\eeq
\beq 
{{\sigma_z}} (n) = \Pi_{j=0}^{n-1} {\mu_x} (j+1/2) ,
\eeq
\beq
 [{\mu}_x (n + 1/2) , {\mu}_z (n^{'} + 1/2) ]= 2 \delta_{n,n^{'} }
\eeq
\beq 
 [{\mu}_z (n + 1/2) , {\mu}_z (n^{'} + 1/2) ]= 0 ,
\eeq
\beq
 [{\mu}_z (n + 1/2) , {\sigma}_x (n^{'} ) ]= 0
\eeq 
The operator ${\mu}_z  (n+1/2 )$ acting on the original spin of the lattice
makes a spin flip of all those spin placed on the left hand side of spin 
at the site n. Therefore ${\mu}_z (n +1/2 ) $ is a kink operator, it introduce
the disorder in the system.
It is very clear from the above analytical relation of the operators that
${\mu}_x$ is related with the allinegment of the spin operator.\\
Here we define the Dirac spinor, 
$ \chi_1 (n) =  \sigma_z (n) \mu_z (n+ 1/2)$ and 
$ \chi_2 (n) =  \sigma_z (n) \mu_z (n- 1/2)$.\\ 
Now our main task is to find the equation of motion for the operators,
${\sigma}_3  (n) $ and ${\mu}_3 (n) $ which help us to build the 
Dirac equation.\\
The equation of motion for the ${\sigma}_z (n)$ is the following:\\
\beq
\frac{\partial \sigma_z (n)}{\partial \tau} = [H, \sigma_z (n)]= {\sigma_x (n) \sigma_z (n) }
\eeq 
The equation of motion for $\mu_z (n+1/2) $ is the following:\\
\bea
\frac{\partial \mu_z (n+1/2)}{\partial \tau} & = & 
\lambda {\mu_x (n +1/2 ) \mu_z (n + 1/2) } \non\\
& & = \lambda \sigma_z (n) \sigma_z (n + 1/2) \mu_z (n + 1/2) \\  
\eea
Now we use the properties
of the $\sigma$ and $\mu$ operators to derive the equation of
motion for the Majorana fields $ \chi_1 (n) $ and $\chi_2 (n) $.\\
\beq
\frac{ \partial \chi_1 (n) }{d \tau} = \frac{\partial \sigma_z (n)}{\partial \tau}
\mu_z (n+ 1/2) + \sigma_z (n) \frac{\partial \mu_z (n)}{\partial \tau}.
\eeq 
\beq
\frac{ \partial \chi_1 (n) }{d \tau} = \sigma_x (n) \sigma_z (n) 
\mu_z (n+ 1/2) + \lambda \sigma_z (n) \sigma_z (n) \sigma_z (n+1) \mu_z (n + 1/2). 
\eeq
\bea
\frac{ \partial \chi_1 (n) }{d \tau} & = & - \sigma_z (n) 
\mu_z (n- 1/2) \mu_z (n+ 1/2) \mu_z (n+ 1/2) \non\\ 
& & + \lambda \sigma_z (n) \sigma_z (n) \sigma_z (n+1) \mu_z (n + 1/2). 
\eea
\beq
\frac{ \partial \chi_1 (n) }{d \tau} = - \chi_2 (n) 
+ \lambda \chi_2 (n + 1). 
\eeq
Now the equations of motion for $\chi_2 (n) $ are
\bea
\frac{ \partial \chi_2 (n) }{d \tau} & = & \frac{\partial \sigma_z (n) }{\partial \tau}
\mu_z (n -1/2) \non\\ 
& & + \sigma_z (n) \frac{\partial \mu_z (n - 1/2) }{\partial \tau} 
\eea
\bea
\frac{ \partial \chi_2 (n) }{d \tau} & = & \sigma_x (n) \sigma_z (n) 
\mu_z (n -1/2) \non\\ 
& & + \lambda \sigma_z (n) \sigma_z (n-1) \sigma_z (n) \mu_z (n - 1/2) 
\eea
\bea
\frac{ \partial \chi_2 (n) }{d \tau} & = & \mu_z (n-1/2) \mu_z (n + 1/2) \sigma_z (n) 
\mu_z (n -1/2) \non\\ 
& & + \lambda \sigma_z (n-1) \mu_z (n - 1/2). 
\eea
After a little bit of calculations and using the relation between the
disorder operators (Eq. 23-30), we finally arrive the equation of motion of $\chi_2 (n) $
as,
\beq
\frac{ \partial \chi_2 (n) }{d \tau} = -\chi_1 (n) 
 + \lambda \chi_1 (n-1). 
\eeq
These two fields, $\chi_1 (n)$ and $\chi_2 (n) $ satisfy the following relations,
$ \{ \chi_1 (n1) , \chi_2 (n2) \} = 2 \delta_{n1, n2} $. One can write down
the above equation in the following compact form, \\
\beq
( {\gamma}^0  \frac{\partial }{\partial t} + {\gamma}^{3} \frac{\partial}{\partial r}
+ m ) \chi =0 . 
\eeq
where ${\chi}^{\dagger} = ( \chi_1 , \chi_2 )$
and $ m= \frac{1 - \lambda}{\alpha} $, 
$ {\gamma}^{0} = \left (\begin{array}{cc}
      0 & 1 \\
    1  & 0
        \end{array} \right ) ,
 {\gamma}^{3} = \left (\begin{array}{cc}
      1 & 0 \\
    0  & -1
        \end{array} \right ) $
.\\ 
Now we prove the presence of Luttinger liquid physics is intrinsic
to the optical cavity array system by comparing the 
analytical relation of Majorana fermion operators with the order
and disorder operator with the free Dirac field in Abelian bosonization
theory.
Here we express the analytical relation between the Majorana
operators and the disorder and Pauli operators:\\
\beq
{\chi}_1 (n) = {\sigma}_z (n) {\mu}_z (n+1/2) = - {\mu}_z (n+ 1/2) {\sigma}_z (n)  
\eeq
\beq
{\chi}_2 (n) = {\sigma}_z (n) {\mu}_z (n-1/2) =  {\mu}_z (n- 1/2) {\sigma}_z (n)  
\eeq
\beq
{\sigma}_z (n) {\chi}_2 (n) =  {\mu}_z (n-1/2) =  {\chi}_2 (n) {\sigma}_z (n)  
\eeq
\beq
{\sigma}_z (n) {\chi}_1 (n) =  {\mu}_z (n+1/2) = - {\chi}_1 (n) {\sigma}_z (n)  
\eeq
\beq
{\sigma}_z (n)=  {\mu}_z (n-1/2) {\chi}_2 (n) =  {\chi}_2 (n) {\mu}_z (n- 1/2)   
\eeq
\beq
{\mu}_z (n+1/2) {\chi}_1 (n) = - {\chi}_1 (n) {\mu}_z (n+1/2)= {\sigma}_z (n)  
\eeq
The above relations can be extended to account for arbitrary space separation between
different operators. Then one obtains the following sets of commutation
relations.\\
\beq
{\sigma}_z ( x1 ) {\mu}_z (x2) = {\mu}_z (x2) {\sigma}_z (x1 ) sign( x1 -x2)  
\eeq
\beq
{\sigma}_z ( x1 ) {\chi} (x2) = {\chi} (x2) {\sigma}_z (x1 ) sign( x1 -x2)  
\eeq
\beq
{\mu}_z ( x1 ) {\chi} (x2) = - {\chi} (x2) {\sigma}_z (x1 ) sign( x1 -x2)  
\eeq
It is very clear from the above analytical relations that
${{\chi}_1}^{\dagger}= {\chi}_1 $ and ${{\chi}_2}^{\dagger} = {{\chi}_2 }$.
The above relation has similarity with the free Dirac field in Abalian
bosonization theory, where Dirac field operator is a local
product of two phase exponential depending on the scalar
field and its dual \cite{gia,gogo}, as one study the
Luttinger liquid physics in Abelian bosonization theory. 
Therefore the Luttinger liquid physics is the
intrinsic to the optical microcavity system.\\
{\bf Conclusions}\\
We have presented an extensive derivation of Dirac equation and the existence 
of Majorana fermion modes for the optical cavity array  
with the relation between Rabi frequency oscillation and the
atom photon coupling strength. We have presented the condition
for massless excitation. We have also presented several analytical relations
between the Majorana field, order and disorder operator.\\
Acknowledgement: The author would like to acknowledge the discussions
with Prof. S. Girvin  
during the international workshop/school on Dirac Materials and 
Chandrashekar lecture at ICTS (December, 2012) and
the library of Raman Research Institute (Mr. Manjunath).
Finally, the author would like to thank Dr. P. K. Mukherjee for reading the
manuscript carefully.

\end{document}